\documentclass[preprint,12pt,3p]{elsarticle}
\usepackage{epsfig,amsfonts,amssymb,latexsym,amsmath,bm,color,array,epsfig}
\biboptions{sort&compress}

\usepackage[active]{srcltx}

\newcommand{\be}{\begin{equation}}
\newcommand{\ee}{\end{equation}}
\newcommand{\bea}{\begin{eqnarray}}
\newcommand{\eea}{\end{eqnarray}}
\newcommand{\beas}{\begin{eqnarray*}}
\newcommand{\eeas}{\end{eqnarray*}}
\newcommand{\ds}{\displaystyle}

\newcommand{\al}{&\!\!}
\newcommand{\Lag}{\mathcal{L}}

\newcommand{\state}[1]{\left|#1\right\rangle}

\newcommand{\X}{X(3872)}
\newcommand{\Jp}{\gamma J/\psi}
\newcommand{\p}{\gamma\psi'}
\newcommand{\pp}{\gamma\psi}

\journal{Physics Letters B}

\begin{document}

\begin{frontmatter}

\title{What can radiative decays of the $X(3872)$ teach us about its nature?}

\author[1]{Feng-Kun Guo}
\author[2]{C. Hanhart}
\author[3]{Yu.S. Kalashnikova}
\author[1,2]{Ulf-G. Mei\ss ner}
\author[3,4,5]{A.V. Nefediev}

\address[1]{Helmholtz-Institut f\"ur Strahlen- und Kernphysik and
Bethe Center for Theoretical Physics, Universit\"{a}t Bonn, D-53115 Bonn, Germany}
\address[2]{Forschungszentrum J\"ulich, Institute for Advanced Simulation, Institut f\"ur Kernphysik and
J\"ulich Center for Hadron Physics, D-52425 J\"ulich, Germany}
\address[3]{Institute for Theoretical and Experimental Physics, B. Cheremushkinskaya 25, 117218 Moscow, Russia}
\address[4]{National Research Nuclear University MEPhI, 115409, Moscow, Russia}
\address[5]{Moscow Institute of Physics and Technology, 141700, Dolgoprudny, Moscow Region, Russia}

\begin{abstract}
Starting from the hypothesis that the $\X$ is a $D\bar D^*$ molecule,
we discuss the radiative decays of the $\X$ into  $\Jp$ and
$\p$ from an effective field theory point of view.
We show that radiative decays are very weakly sensitive to the long-range structure of the $\X$.
In particular, contrary to earlier claims,
we argue that the experimentally determined ratio of the mentioned
branching fractions is not in conflict with a wave
function of  the $\X$ that is dominated by the $D\bar D^*$
hadronic molecular component.
\end{abstract}


\end{frontmatter}

\section{Introduction}

The $\X$ was discovered by the Belle Collaboration in 2003~\cite{Choi:2003ue}.
It has a mass extremely close to the $D^0\bar D^{*0}$ threshold, and thus it
has been regarded as one of the most promising candidates for a hadronic
molecule, which can be either an $S$-wave bound
state~\cite{Voloshin:1976ap,De Rujula:1976qd,Tornqvist:2004qy,tornqvist,swanson,wong} or a virtual
state in the  $ D\bar D^*$ system~\cite{Hanhart:2007yq}.
Its quantum numbers were determined by the
LHCb Collaboration to be $J^{PC}=1^{++}$~\cite{Aaij:2013zoa} 10 years after the discovery.

Other models exist in addition to the hadronic molecule interpretation,
which include a radial excitation of the $P$-wave charmonium
$\chi_{c1}(2P)$~\cite{Barnes:2005pb},
a tetraquark~\cite{polosa}, a mixture of an ordinary charmonium
and a hadronic
molecule~\cite{close,suzuki}, or a state generated in the coupled-channel dynamical scheme ~\cite{YuSK,Danilkin:2010cc}.
It was claimed in Ref.~\cite{Swanson:2004pp} that the radiative decays of the
$\X$ into the $\Jp$ and $\p$ (here and in what follows $\psi'$ denotes
$\psi(2S)$) are very sensitive to its
structure. Especially, using vector meson dominance and  a quark model, in
Ref.~\cite{Swanson:2004pp} it was predicted that the ratio
\begin{equation}
R \equiv \frac{\mathcal{B}(\X\to \p) }{\mathcal{B}(\X\to \Jp) }
\label{eq:ratio}
\end{equation}
is about $4\times10^{-3}$, if the $\X$ is a hadronic molecule with the dominant
component $D^0\bar D^{*0}$ plus a small admixture of the $\rho J/\psi$ and
$\omega J/\psi$. Various quark model calculations predict this ratio in a
range as wide as from approximately 0.6 and up to about 6 assuming a
$c\bar c$ nature for the $\X$, see a few paradigmatic examples collected in
Table~\ref{widths}.
Such a large uncertainty in the predictions as well as the fact that
the values $R>1$ are preferred should not come as a
surprise, since the quark model assignment for the $\X$ is that of a
radially excited $\chi_{c1}(2P)$ $c\bar{c}$ state
while a radiative decay matrix element is proportional to the overlap
integral of the initial state and the final state
wave functions. Thus, on the one hand, such an overlap is very
sensitive to the details of the wave functions, in particular
to the position of their nodes. On the other hand, the overlap of the
radially excited $\chi_{c1}(2P)$ charmonium
wave function with the one-node $\psi'$ wave function is expected
to be larger than its overlap with the nodeless
$J/\psi$ wave function.

The ratio $R$ was first measured by the BABAR Collaboration with the result
$3.4\pm1.4$~\cite{Aubert:2008ae}. Later on, the Belle Collaboration reported a
negative result for the $\p$ mode, and set the upper limit
$R<2.1$ at $90\%$ confidence level~\cite{Bhardwaj:2011dj}. Very recently,
the LHCb Collaboration reported the ratio~\cite{Aaij:2014ala}
\be
R=2.46\pm0.64\pm0.29,
\label{RLHCb}
\ee
which is consistent with both previous measurements.
Based on the mentioned claim of Ref.~\cite{Swanson:2004pp} this result
was interpreted as a strong evidence that the $\X$ cannot be a hadronic
molecule.
However, it was found in Ref.~\cite{dong}, which updates earlier
calculations in Refs.~\cite{Dong:2008gb,Dong:2009yp}, in a phenomenological
study allowing for both a molecular as well as a compact component of the $\X$
that an enhanced decay of the $\X$ into $\p$ compared to $\gamma J/\psi$ is
fully compatible with a predominantly molecular nature of $\X$. An admixture of
5--12\% of a $c\bar c$ component was sufficient to explain the data.
In this paper we critically re-investigate the validity of the claim of
Ref.~\cite{Swanson:2004pp} from an effective field theory point of view. In
particular we demonstrate that, contrary to earlier claims, radiative decays do
not allow one to draw conclusions on the nature of $\X$ and therefore confirm
qualitatively the findings  of Ref.~\cite{dong} that the observed ratio is not
in conflict with a predominantly molecular nature of the $\X$.

\begin{table}[t]
\begin{center}
\begin{tabular}{|l|c|c|c|}
\hline
&$\Gamma(X\to \Jp)$ [keV] &$\Gamma(X\to \p)$ [keV] &$R$\\
\hline
Ref.~\cite{BG}&11&64&5.8\\
\hline
Ref.~\cite{Barnes:2005pb}&70&180&2.6\\
\hline
Ref.~\cite{Badalian:2012jz}&50-70&50-60&$0.8\pm 0.2$\\
\hline
\end{tabular}
\end{center}
\caption{Some paradigmatic examples of quark-model estimates for
the radiative decays of the $2^3P_1$ charmonium.}
\label{widths}
\end{table}

\section{Generalities}

According to Ref.~\cite{weinberg} one may define the molecular component of a
bound state by the probability to find the continuum component in the wave
function of the physical state. This definition provides a close link between
the significance of hadronic loops and hadronic molecules.
In studies of quarkonia, the importance of hadronic loops is case dependent. In
some transitions as discussed in, for example,
Refs.~\cite{Guo:2009wr,Guo:2010ak,Guo:2010ca}, they are expected to give
sizeable contributions.
In contradistinction hadronic molecules have
large effective coupling constants to the continuum as follows straightforwardly
from the analysis of Ref.~\cite{weinberg}, and a pure molecule only couples to
its constituents.
As a consequence, in their decays hadronic loops are by definition a leading
order effect. Because of this, hadronic molecules leave unique imprints in some
properly chosen observables, but not in
all: as we discuss in this paper in detail, in order to quantitatively control
(ratios of) transition rates, additional information, not at all linked
to the nature of the state under investigation, on the matrix element
that connects the continuum
state to the final state might be necessary. In addition, not all observables
are related to the long-range tail of the wave function of a molecular state. In
particular, we demonstrate that the radiative decays are much more sensitive to the
short-range parts of the $\X$ wave function rather than to the long-distance nature of the $\X$.

The situation is analogous to that of the $D_{s0}^*(2317)$: when being treated
as a $c\bar s$ state meson loops appear in the effective field theory only at
subleading orders and give a small contribution to the
decays~\cite{mehenspringer}. On the other hand, if the assumed structure is a
$DK$ molecule, meson loops are a leading order effect~\cite{Dsdecays}. However,
this does not imply that all observables allow one to distinguish between the
two scenarios: in Ref.~\cite{Dsdecays} it was argued that while the strong
decays are sensitive to the nature of the state the radiative decays are not
because there are short-range contributions present already at the leading
order.

\begin{figure*}[t]
\begin{center}
\begin{tabular}{ccc}
\raisebox{13mm}{(a)}~\epsfig{file=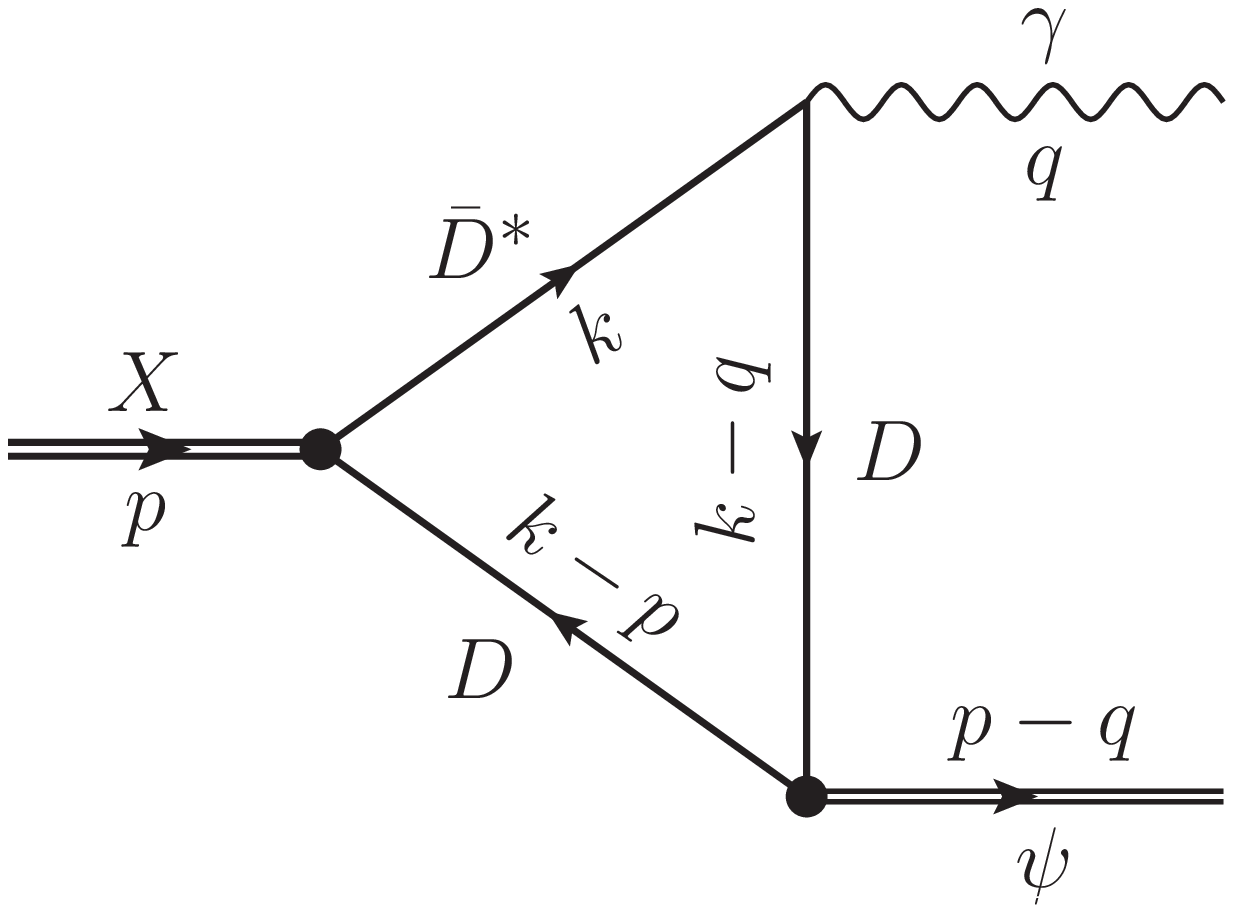,width=0.25\textwidth}&
\raisebox{13mm}{(b)} ~\epsfig{file=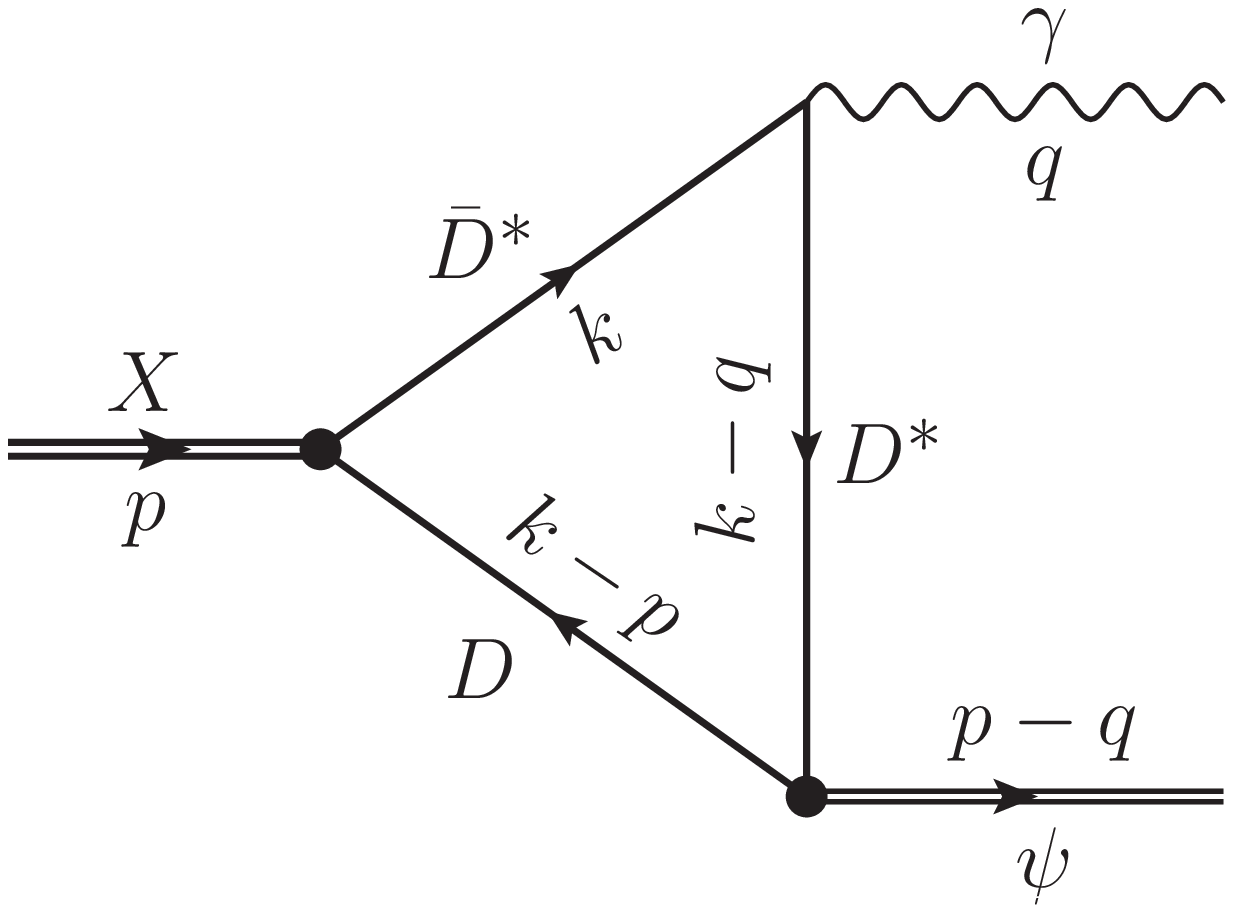,width=0.25\textwidth}&
\raisebox{13mm}{(c)} ~\epsfig{file=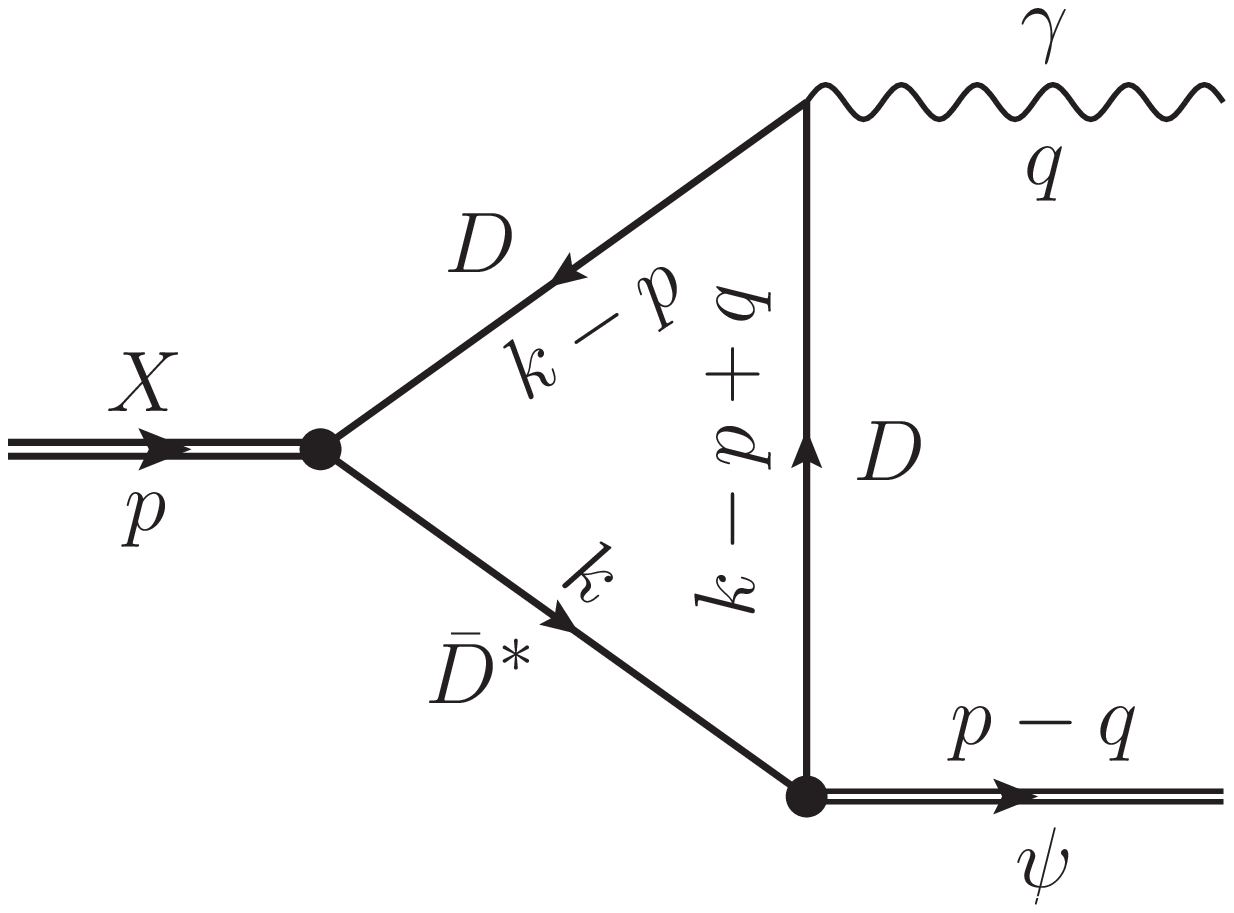,width=0.25\textwidth}\\
\raisebox{13mm}{(d)} ~\epsfig{file=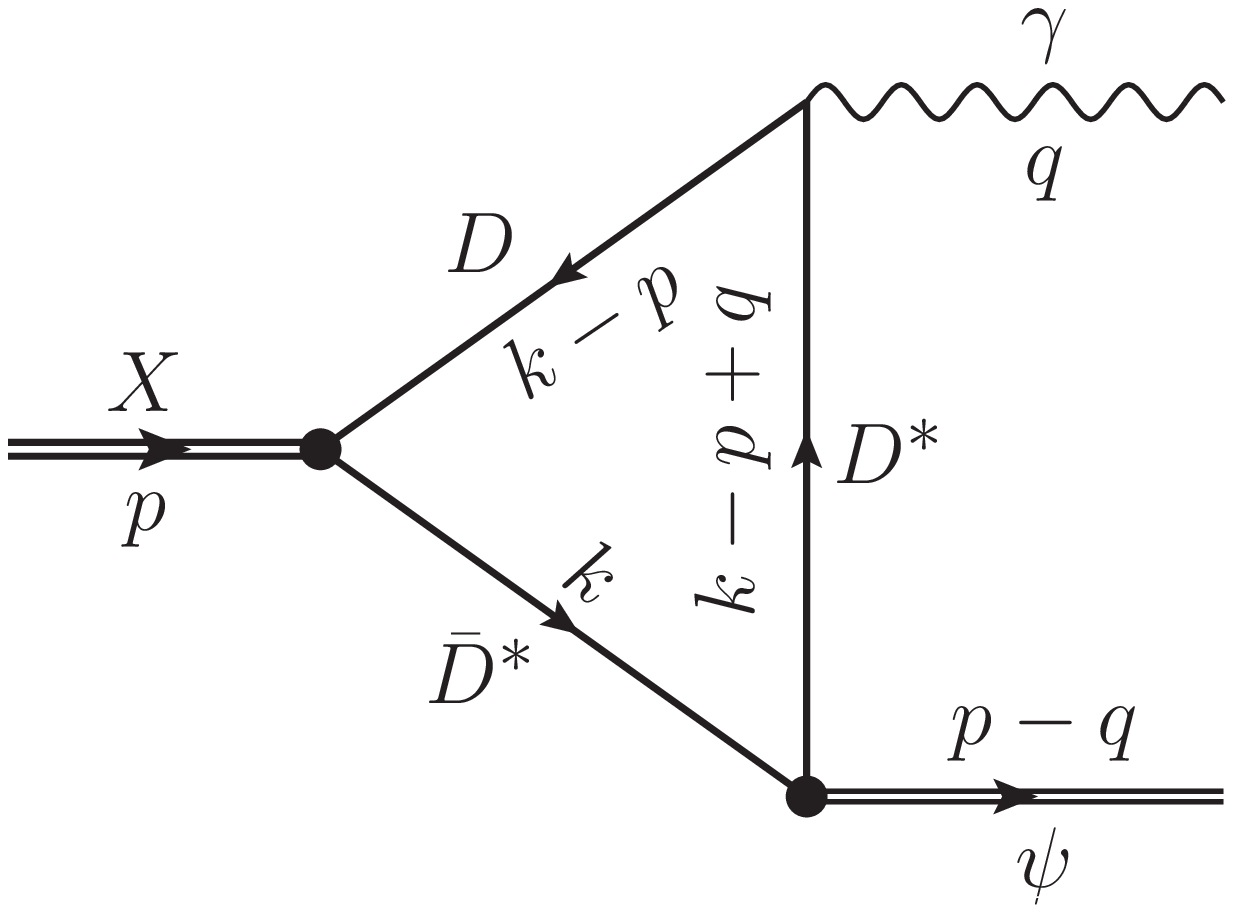,width=0.25\textwidth}&
\raisebox{13mm}{(e)} ~\raisebox{2.5mm}{\epsfig{file=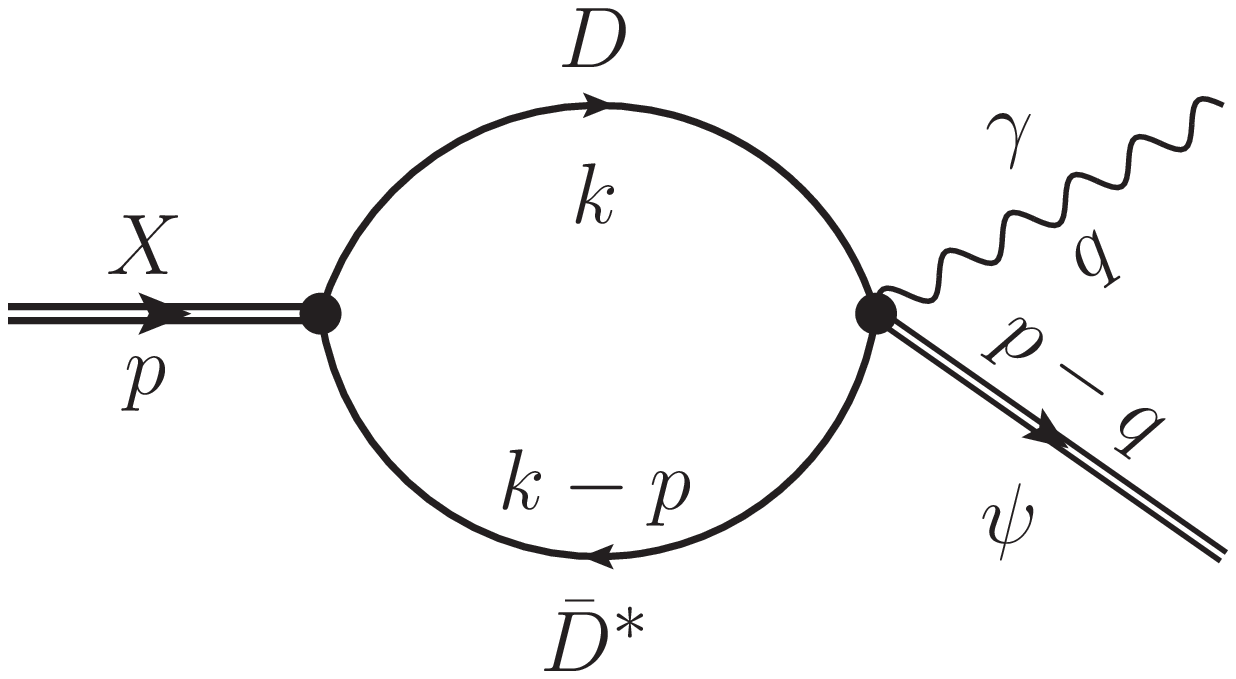,width=0.25\textwidth}}&
\raisebox{13mm}{(f)} ~\raisebox{5.5mm}{\epsfig{file=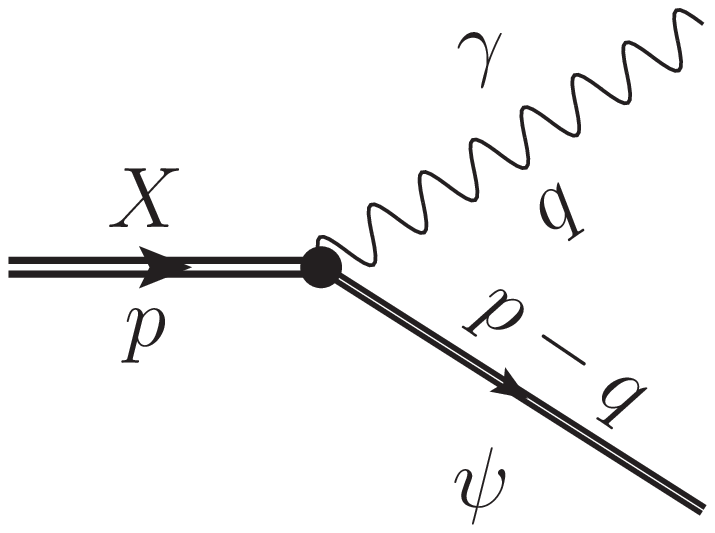,width=0.15\textwidth}}
\end{tabular}
\caption{Decay mechanism for the transitions $\X\to \pp$ where
$\psi=J/\psi$ or $\psi'$ if the $\X$ is a $D\bar D^*$ hadronic molecule. Here
both charged and neutral charmed mesons are taken into account in the first
four diagrams. The charge
conjugated diagrams are not shown but included in the calculations.}
\label{fig:triangle}
\end{center}
\end{figure*}

The decay mechanisms for  $\X$ into the $\pp$, with $\psi$ denoting $J/\psi$ or
$\psi'$, are shown in Fig.~\ref{fig:triangle}. The charge conjugated diagrams
are not depicted but are taken into account in the decay amplitude.
We use the diagrams shown in Fig.~\ref{fig:triangle} to calculate the $\X$
radiative decay widths  employing  a covariant approach. The details
are presented
in the next section.
Heavy quark spin symmetry (HQSS) is used wherever appropriate to relate the
vector and pseudoscalar charmed mesons. Our phase convention for the charge
conjugation of the charmed mesons is
\begin{equation}
\mathcal{C} D^{(*)} \mathcal{C}^{-1} = \bar D^{(*)}.
\end{equation}
Under this convention, the wave function of the $\X$ as a pure hadronic
molecule may be written as
\begin{equation}
\state{\X} = \frac1{\sqrt{2} } \left( \state{D\bar D^*} + \state{\bar D D^*} \right).
\end{equation}

\section{Formalism}

The loop amplitude for the decay $X_\sigma(p)\to\gamma_\lambda(q)\psi_\mu(p')$
($p'=p-q$) is given by the sum of the diagrams (a)-(e) depicted in
Fig.~\ref{fig:triangle}.

The $X_\sigma(p)\to D\bar{D}^*_\nu(k)$ interaction Lagrangian reads
\be
\Lag_X=\frac{x_0}{\sqrt{2}}X_\sigma^{\dag}\left(D^{*0\,\sigma}\bar{D}^0+D^0\bar{D}^{*0\,\sigma}\right)
+\frac{x_c}{\sqrt{2}}X_\sigma^{\dag}\left(D^{*+\,\sigma}D^-+D^+D^{*-\,\sigma}\right)+\mbox{h.c.},
\label{lagXnr}
\ee
where the values of the coupling constants of the $\X$ to the charged and
neutral charmed mesons are similar, see, for example,
Ref.~\cite{Gamermann:2009fv,Guo:2014hqa}, so in what follows we do not
distinguish between them and set $x_c =x_0 =x$. For convenience, we relate
the relativistic coupling to the nonrelativistic one,
\be
x=x_{\rm nr}\sqrt{m_Xm_*m},
\ee
where $m$, $m_*$, and $m_X$ are the $D$-meson, $D^*$-meson, and the $\X$ mass, respectively. We do not distinguish
between the neutral and charged $D$- and $D^*$-meson masses.
The value of $x^{\rm  nr}$ was extracted from the binding energy of the
$X(3872)$ in the hadronic molecule picture in Ref.~\cite{Guo:2013zbw}.
Then the $X_\sigma(p)\to D\bar{D}^*_\nu(k)$ vertex takes the form
\be
\Gamma_{\sigma\nu}^{(X)}(p,k)=\frac{1}{\sqrt{2}}x_{\rm nr}\sqrt{m_Xm_*m}\,g_{\sigma\nu}.
\ee

The vertices relating the vector field $\psi$ ($\psi=J/\psi,\psi'$) to the $D$
and $D^*$ mesons follow from the
Lagrangian
\bea
\Lag_\psi \al=\al
i \,
g_{\bar{D}D}(\bar{D}\overset{\leftrightarrow}{\partial_\mu}D)\psi^{\mu\dag}
-i\, g_{\bar{D}^*D}\varepsilon_{\mu\nu\alpha\beta}
\left[(\partial^\alpha\bar{D}^*_\nu)(\partial^\beta D)-
\left(\partial^\beta \bar D\right) (\partial^\alpha{D}^*_\nu)
\right]\psi^{\mu\dag} \nonumber\\
\al\al
- i \,
g_{\bar{D}^*D^*}
\left(\bar{D}^*_\nu\overset{\leftrightarrow}{\partial_\mu}D^{*\nu}+
(\partial_\nu\bar{D}^*_\mu)D^{*\nu}-\bar{D}^{*\nu}(\partial_\nu D^*_\mu)
\right)\psi^{\mu\dag} +\mbox{h.c.},
\label{LpsiDD}
\eea
where the couplings are related via the heavy-quark symmetry.
With the help of the nonrelativistic Lagrangians from
Refs.~\cite{Colangelo:2003sa,Guo:2009wr,Guo:2010ak} one finds
\be
g_{\bar{D}D}=g_2m\sqrt{m_\psi},\quad g_{\bar{D}^*D}=2g_2\sqrt{\frac{m\, m_\psi}{m_*} },
\quad g_{\bar{D}^*D^*}=g_2m_*\sqrt{m_\psi},
\label{eq:couplings}
\ee
with the nonrelativistic coupling constant $g_2$ used in
Refs.~\cite{Guo:2009wr,Guo:2010ak}.
Thus, the $\psi_{\mu}(p)\to\bar{D}(k_1)D(-k_2)$, $\psi_{\mu}(p)\to \bar{D}^*_{\nu}(k_1)D(-k_2)$, and
$\psi_{\mu}(p)\to \bar{D}^*_\alpha(k_1)D^*_\beta(-k_2)$ vertices read, respectively,
\be
V_\mu^{\bar{D}D}(k_1,-k_2)=g_2\sqrt{m_\psi}m (k_1+k_2)_\mu,
\label{psidd}
\ee
\be
V_{\mu\nu}^{\bar{D}D^*}(k_1,-k_2)=2g_2\sqrt{\frac{m_\psi m}{m_*}}\epsilon_{\mu\nu\alpha\beta}k_2^\alpha k_1^\beta,
\label{psiddstar}
\ee
\be
V_{\mu\alpha\beta}^{\bar{D}^*D^*}(k_1,-k_2)=g_2\sqrt{m_\psi}m_*\Bigl[(k_1+k_2)_{\mu}g_
{\alpha\beta}-(k_1+k_2)_{\beta}g_{\mu\alpha}-
(k_1+k_2)_{\alpha}g_{\mu\beta}\Bigr].
\label{psidsds}
\ee

Next, we need the couplings of the photon to the open-charm states.
The leading electric couplings emerge from gauging the kinetic terms
for the charged heavy mesons. The importance of
the charged component of the $\X$ wave function for the radiative
decays was stressed in Ref.~\cite{Aceti:2012cb}.
The $D^\pm(k_1) \to D^\pm(k_2)\gamma_{\lambda}(q)$ ($k_1=k_2+q$) electric
vertex reads
\be
\Gamma^{(e)}_\lambda(k_1,k_2)=e(k_1+k_2)_{\lambda},\quad q=k_1-k_2,
\label{G1}
\ee
where $e$ is the electric charge.
The $D^{*\pm}_{\mu}(k_1) \to D^{*\pm}_{\nu}(k_2)\gamma_{\lambda}(q)$ ($k_1=k_2+q$) one reads
\be
\Gamma^{(e)}_{\mu\nu\lambda}(k_1,k_2)=e\Bigl[(k_1+k_2)_{\lambda}g_{\mu\nu}-k_{1\nu}g_{\mu\lambda}-
k_{2\mu}g_{\nu\lambda}\Bigr].
\label{psipsigam}
\ee
Both vertices satisfy the appropriate Ward identities,
\bea
&q^\lambda\Gamma^{(e)}_\lambda(k_1,k_2)=e\Bigl[S^{-1}(k_2)-S^{-1}(k_1)\Bigr],&
\nonumber\\[-2mm]
\label{WI}\\[-2mm]
&q^\lambda\Gamma^{(e)}_{\mu\nu\lambda}(k_1,k_2)=e\Bigl[(S^{-1}(k_2))_{\mu\nu}-(S^{-1}(k_1))_{\mu\nu}\Bigr],&
\nonumber
\eea
where the $D$-meson propagator is
\be
S(p)=\frac{1}{p^2-m^2+i\varepsilon},
\ee
and the $D^*$ propagator and its inverse form are
\be
S_{\mu\nu}(p)=\frac{1}{p^2-m_*^2+i\varepsilon}\left(-g_{\mu\nu}+\frac{p_{\mu}p_{\nu}}{m_*^2}\right),
\quad
(S^{-1}(p))_{\mu\nu}=-(p^2-m_*^2)g_{\mu\nu}+p_{\mu}p_{\nu}.
\label{SS}
\ee

Finally, the vertex (\ref{psiddstar}) gives rise to
a four-point vertex $D\bar{D}^*\psi \gamma$ after gauging, see
diagram (e) in Fig.~\ref{fig:triangle}.

To extract the magnetic vertices we use the covariant generalisation
of the nonrelativistic Lagrangian
in Refs.~\cite{Amundson:1992yp,Cheng:1992xi}, which reads
\bea
&&\Lag_m=iem_*F_{\mu\nu} D_a^{*\mu\,\dag}\left(\beta
Q_{ab}-\frac{Q_c}{m_c}\delta_{ab}\right)D^{*\nu}_b \nonumber\\
&&\hspace*{30mm}+e\sqrt{mm_*}\epsilon_{\lambda\mu\alpha\beta}v^\alpha\partial^\beta
A^\lambda\left[D_a^{*\mu\,\dag}\left(\beta
Q_{ab}+\frac{Q_c}{m_c}\delta_{ab}\right)D_b+\mbox{h.c.}\right],
\label{Lagm}
\eea
where $v^\mu$ is the four-velocity of the heavy quark with
$v^\mu v_\mu =1$, $Q=\text{diag}(2/3,-1/3)$ is the light
quark charge matrix, and $m_c$ and $Q_c$ are the charmed quark
mass and its charge, $Q_c=2/3$, respectively. In the above
Lagrangian, the terms proportional to $Q_c/m_c$ come from the magnetic moment
of the charm quark, and the $\beta$-terms are from the nonperturbative
light-flavour cloud in the charmed meson.

Then the magnetic $D^{*a}_{\mu}(k_1) \to D^{*b}_{\nu}(k_2)\gamma_{\lambda}(q)$
and $D^{*a}_{\mu}(k_1) \to
D^b(k_2)\gamma_{\lambda}(q)$ ($k_1=k_2+q$) vertices read
\be
\Gamma_{\mu\nu\lambda}^{(m)ab}(q)=em_*(q_\nu g_{\mu\lambda}-q_\mu g_{\nu\lambda})\left(\beta
Q_{ab}-\frac{Q_c}{m_c}\delta_{ab}\right)
\label{DsDsgamm}
\ee
and
\be
\Gamma_{\mu\lambda}^{(m)ab}(q)=e\sqrt{mm_*}\varepsilon_{\mu\lambda\alpha\beta}v^{\alpha}q^\beta \left(\beta
Q_{ab}+\frac{Q_c}{m_c}\delta_{ab}\right),
\label{DsDgamm}
\ee
respectively. Notice that both vertices (\ref{DsDsgamm}) and
(\ref{DsDgamm}) are manifestly transversal with respect to the photon
momentum $q^\lambda$.

With the given ingredients the expression for the loop amplitude
of the radiative decay $X\to\gamma\psi$ reads
\be
M^{\rm loop}=\varepsilon^\mu(\psi)\varepsilon^\sigma(X)\varepsilon^\lambda(\gamma)M_{\mu\sigma\lambda}^{\rm
loop},
\ee
where
\be
M_{\mu\sigma\lambda}^{\rm loop}=\frac{1}{\sqrt{2}}exg_2m\sqrt{m_Xm_\psi}
\int\frac{d^4k}{(2\pi)^4}S_\sigma^\nu(k)S(k-p)J_{\mu\nu\lambda}(k).
\label{Mampl}
\ee

The tensor $J_{\mu\nu\lambda}(k)$ acquires contributions from
diagram (a)--(e) in Fig.~\ref{fig:triangle},
\be
J_{\mu\nu\lambda}(k)=J_{\mu\nu\lambda}^{(a)m}(k)+J_{\mu\nu\lambda}^{(b)e}(k)+J_{\mu\nu\lambda}^{(b)m}(k)
+J_{\mu\nu\lambda}^{(c)e}(k)+J_{\mu\nu\lambda}^{(d)m}(k)+J_{\mu\nu\lambda}^{(e)e}(k),
\label{J}
\ee
where the superscripts $e$ and $m$ label the electric and magnetic
contributions, respectively. The individual contributions read
\bea
&&\ds J_{\mu\nu\lambda}^{(a)m}(k)=\frac13
m\left(\beta+\frac{4}{m_c}\right)\varepsilon_{\nu\lambda\alpha\beta}p^\alpha
q^\beta\frac{(2k-p-q)_\mu}{(k-q)^2-m^2},\\[1mm]
&&\ds J_{\mu\nu\lambda}^{(b)e}(k)=2\varepsilon_{\mu\rho\alpha\beta}\frac{(k-p)^\alpha(k-q)^\beta}
{(k-q)^2-m_*^2}
[(2k-q)_\lambda g^\rho_\nu-(k-q)_\nu g^\rho_\lambda-k^\rho g_{\nu\lambda}],\\[1mm]
&&\ds J_{\mu\nu\lambda}^{(b)m}(k)=\frac23m_*\left(\beta-\frac{4}{m_c}\right)\varepsilon_{\mu\rho\alpha\beta}\frac{
(k-p)^\alpha(k-q)^\beta}
{(k-q)^2-m_*^2}[q_\nu g^\rho_\lambda-q^\rho g_{\nu\lambda}],\\[1mm]
&&\ds J_{\mu\nu\lambda}^{(c)e}(k)=2\varepsilon_{\mu\nu\alpha\beta}(k-p+q)^\alpha
k^\beta\frac{(2k-2p+q)_\lambda}{(k-p+q)^2-m^2},\\
&&\ds J_{\mu\nu\lambda}^{(d)m}(k)=\frac13m_*\left(\beta+\frac{4}{m_c}\right)[(2k-p+q)_\mu
g_{\beta\nu}-(2k-p+q)_\beta g_{\mu\nu}-(2k-p+q)_\nu g_{\beta\mu}]\nonumber\\
&&\ds \hspace*{40mm}\times\frac{\varepsilon_{\alpha\lambda\gamma\delta}p^\gamma q^\delta
}{(k-p+q)^2-m_*^2}\left(-g^{\alpha\beta}+\frac{(k-p+q)^\alpha(k-p+q)^\beta}{m_*^2}\right),\\[1mm]
&&\ds J_{\mu\nu\lambda}^{(e)e}(k)=-2\varepsilon_{\mu\nu\lambda\alpha}p^\alpha.
\eea
In the expressions above the heavy-quark four-velocity is substituted
by the $\X$ four-velocity and the contribution of the
conjugated loops is taken into account explicitly.
The amplitude (\ref{Mampl}) is gauge invariant. This follows from the
transversality of the magnetic
vertices of Eqs.~(\ref{DsDsgamm}) and (\ref{DsDgamm})
as well as from the Ward identities of Eq.~(\ref{WI}).

It is easy to verify that the loop integral in the amplitude (\ref{Mampl}) is divergent.
Therefore, to render the result well defined, one needs to
include in addition to the loop amplitude (\ref{Mampl}) described above
the $X\gamma\psi$ counterterm amplitude (diagram (f) in Fig.~\ref{fig:triangle}),
\be
M^{\rm cont}=
\lambda\varepsilon_{\mu\sigma\lambda\nu}\varepsilon^\mu(\psi)\varepsilon^\sigma(X)\varepsilon^\lambda(\gamma)q^\nu,
\label{Mcont}
\ee
which is also manifestly gauge invariant. The strength of the contact
interaction $\lambda$ is subject to
renormalisation to absorb the divergence of the loops, so that the contact
amplitude (\ref{Mcont}) with the renormalised strength $\lambda_r$
provides a finite contribution to the total decay width.
In the next section two different ways are presented on how to
estimate the size of this finite contribution. The necessity to
include a contact term at leading order shows that the radiative decays
are not only sensitive to the long-range parts of the matrix element entering via the loops but also to the short-range
structure of the wave function which is not known. Because of that, as a matter of principle,
the radiative decays of $\X$ cannot be used as a source of information on its long-distance structure.

In the calculations we used the following values for the masses~\cite{PDG}:
$$
m=1865~\mbox{MeV},~m_*=2007~\mbox{MeV},~
m_X=3872~\mbox{MeV},~m_{J/\psi}=3097~\mbox{MeV},~m_{\psi'}=3686~\mbox{MeV}.
$$

For the magnetic coupling of the charmed mesons, the
parameters are~\cite{Hu:2005gf}
\be
\beta^{-1}=379~\mbox{MeV},\quad m_c=1876~\mbox{MeV}.
\ee

The value of the coupling constant $x_{\rm nr}$ is very uncertain because the
value of its mass, or the binding energy, is not known precisely. In
Ref.~\cite{Guo:2013zbw}, it was extracted to be $|x_{\rm nr}| =
0.97^{+0.40}_{-0.97}$~GeV$^{-1/2}$. The coupling constants for $J/\psi$ and $\psi'$
to the charmed mesons, $g_2$ and $g_2'$, cannot be measured directly and are
badly known. We observe, see Eq.~(\ref{Mampl}), that the width in the
pure molecular picture is
proportional to $|x_{\rm nr} g_2|^2$. In order to give definite values for the
partial widths, we set the finite part of the counterterms to zero,
$\lambda_r^{(\prime)}=0$, and define ratios
\be
r_x\equiv\left|\frac{x_{\rm nr}}{x_{\rm nr}^{(0)}}\right|,\quad
r_g\equiv\left|\frac{g_2}{g_2^{(0)}}\right|,\quad
r_g'\equiv\left|\frac{g_2'}{g_2^{(0)}}\right|,
\label{eq:rcouplings}
\ee
where $|x_{\rm nr}^{(0)}|=0.97$~GeV$^{-1/2}$~\cite{Guo:2013zbw} and
$|g_2^{(0)}|=2$~GeV$^{-3/2}$ is taken from model
estimates~\cite{Colangelo:2003sa,Guo:2010ak}.

The integrals are evaluated using  dimension regularisation with the
$\overline{\rm MS}$ subtraction scheme at the scale $\mu=m_X$.
Numerical calculations are  performed with the help of the FeynCalc
\cite{FeynCalc} and the LoopTools \cite{LoopTool} packages for Mathematica.

\begin{table}[t]
\begin{center}
\begin{tabular}{|l|c|c|c|}
\hline
                                 &$\mu=m_X/2$&$\mu=m_X$&$\mu=2m_X$\\
\hline
$\Gamma(X\to\gamma J/\psi)$ [keV] &9.7$(r_xr_g)^2$        &23.5$(r_xr_g)^2$    &43.2$(r_xr_g)^2$      \\
\hline
$\Gamma(X\to\gamma\psi')$ [keV]   &3.8$(r_xr_g')^2$        &4.9$(r_xr_g')^2$      &6.0$(r_xr_g')^2$       \\
\hline
$\ds R=\frac{\Gamma(X\to\gamma\psi')}{\Gamma(X\to\gamma J/\psi)}$ &0.39$(g_2'/g_2)^2$ &0.21$(g_2'/g_2)^2$
&0.14$(g_2'/g_2)^2$    \\
\hline
\end{tabular}
\end{center}
\caption{The calculated radiative decay widths $\Gamma(X\to\gamma\psi)$ for
$\psi=J/\psi,\psi'$ and their ratio $R$. Here, $g_2 (g_2')$ are the spin
symmetric coupling constants of the $J/\psi(\psi')$ to the charm
meson--antimeson pair, see Eq.~\eqref{eq:couplings}, $r_x$ and $r_g^{(\prime)}$
are defined in Eq.~\eqref{eq:rcouplings}. }
\label{tab:res}
\end{table}

\section{Results and discussion}

Our numerical results for the partial radiative decay widths for the $\X$ are
displayed in the middle column of Table~\ref{tab:res}. To arrive at these
results, the contact terms were set to zero ($\lambda_r^{(\prime)}=0$) and the couplings
were chosen as explained in the previous section.

Although the badly known constant $x_{\rm nr}$ responsible for the $\X$ coupling to
the charmed mesons drops from the ratio $R$, still additional assumptions
are necessary in order to connect the results from Table~\ref{tab:res}
to the actual data, namely (i) to fix the ratio of the coupling constants of the $\psi'$ and $J/\psi$,
$g_2'/g_2$, which has nothing to do with the nature of the X(3872), and (ii) to estimate the size of the contact
interaction contribution to the width, which is only sensitive to the short-range structure of the $X$.

First one needs to fix the ratio $g_2'/g_2$. If both couplings were equal and $\lambda_r^{(\prime)}=0$, then
indeed the $\gamma\psi'$ channel would be suppressed, although a lot less than
claimed in Ref.~\cite{Swanson:2004pp}. However, already a value of
$g_2'/g_2\sim 3$ lying in a natural range is sufficient to bring $R$ in accordance with the experiment
also for a purely molecular $\X$, see Eq.~(\ref{RLHCb}). In this context it is
interesting to observe that Ref.~\cite{dong} finds $g_2'/g_2\sim 2$ referring to
the analysis presented in Ref.~\cite{Colangelo:2003sa}.

Second, so far the counterterms were set to zero. From the information we have
available there is no way to fix their strength $\lambda_r^{(\prime)}$ (we do not agree to
the claim of Ref.~\cite{Aceti:2012cb} that the strength can be taken directly
from a model of the formation of the $\X$, since different scales are involved
in scattering and decay).
One way to estimate the size of the contact term contribution to the width is to
vary the regularisation scale used in the evaluation of the loops.
The impact of a variation of $\mu$ from $m_X/2$ to values as large as $2m_X$ is
also presented in Table~\ref{tab:res}.
Since any physical amplitude should be independent of $\mu$ the variation
displayed in Table~\ref{tab:res} should be compensated by a corresponding
variation in the counterterm. In this sense the observed variation in the width
is a measure of the size of the counterterm. This confirms the claim made
earlier in this paper that for the radiative decays of the $\X$ short-range
contributions are of similar importance as their long-range counter parts.

Alternatively one could estimate the size of the counterterms by employing a
model. Indeed, since the counterterms parametrise short-range physics they may
be modelled by a heavy quark loop (cf. discussion in Ref.~\cite{dong}). We
consider a few paradigmatic examples of the estimates for the radiative decay
widths of the $2^3P_1$ $c\bar{c}$ charmonium found in the literature, which are
collected in Table~\ref{widths}.
These estimates for the $\Jp$ mode appear to be 2-3 times larger than the result
quoted in Table~\ref{tab:res} if the ratios defined in Eq.~\eqref{eq:rcouplings}
are taken to be unity, while for the $\p$ mode they exceed the molecule estimate
by more than one order of magnitude. Unfortunately, it is very difficult to estimate the uncertainties
of the results of quark models. Thus, as an anchor, we take the averaged values
$\Gamma(X(c\bar{c})\to \Jp)\simeq 50$~keV and $\Gamma(X(c\bar{c})\to
\p)\simeq100$~keV with the ratio $R\simeq 2$.
However, these numbers cannot be used directly as an estimate for the
counterterm contributions since both the size of the hadron loops presented in
Table~\ref{tab:res} as well as the size of the quark loops presented in
Table~\ref{widths} are based on the condition that the normalisation of the wave
function is saturated by the hadronic loops or the quark loop, respectively.
Therefore, in order to add both results one needs to multiply the widths from
the hadronic loops by $(1-Z)$ and those from the quark loop by $Z$, where $Z$
denotes the probability to find the compact component in the physical wave
function of the $\X$~\cite{weinberg}\footnote{It was shown in
Ref.~\cite{evidence} that inelastic channels do not destroy this logic.}.
Accordingly, $Z=0$ refers to a pure molecule while $Z=1$ points to a purely
elementary state.
Then $Z\sim 0.1$ brings the quark loop contribution to the same order of
magnitude with that of the hadron loops, which clearly allows one to fit the
data.
This observation is in line with that of Ref.~\cite{dong}.
Specifically, for $Z\sim 0.3$-$0.4$ found in Ref.~\cite{Kalashnikova:2009gt} from
the combined data analysis on the $D^0\bar{D}^0\pi^0$ and $\pi^+\pi^- J/\psi$
decay modes of the $X$, the ratio $g_2'/g_2$ needed to bring $R$ in accordance
with the experiment is reduced to approximately 2 in line with the
ratio of couplings found in Ref.~\cite{dong}  referring to
the analysis of Ref.~\cite{Colangelo:2003sa}.

It should be stressed, however, that these findings have to be interpreted with
caution, not only since the numbers presented in the tables are highly
uncertain: as a matter of principle it is not possible to identify in a hadronic
effective field theory the physics of the short-range contributions.
The latter could as well be, for example, higher momentum components of the
hadronic wave function. At this point all one can conclude is that the radiative
decays of the $\X$, and especially their ratio $R$, is very weakly sensitive to the long-range
structure of the $X$, and thus they cannot be used to rule out the picture that the $\X$ is dominantly a hadronic
molecule. In
order to make statements on the hadronic molecule structure of the $X$, one
needs to measure its decays which are sensitive to the long-distance physics,
such as $X\to D\bar D\gamma$ or $X\to D\bar D\pi$, see, for example,
Refs.~\cite{Voloshin:2003nt,Fleming:2007rp,Hanhart:2007yq,Guo:2014hqa,Kalashnikova:2009gt,Stapleton:2009ey}.
\bigskip

The authors would like to thank F. Llanes--Estrada for a careful reading the manuscript and for valuable comments.
This work is supported in part by the DFG and the NSFC through funds provided
to the Sino-German CRC 110 ``Symmetries and the Emergence of Structure in QCD''
(NSFC Grant No. 11261130311), by the EU Integrated Infrastructure Initiative
HadronPhysics3 (Grant No. 283286), by the Russian presidential programme
for the support of the leading scientific schools (Grant No. NSh-3830.2014.2), and by NSFC (Grant No.
11165005).

\end{document}